\pdfoutput=1
\documentclass[12pt]{article}
\usepackage[english]{babel}
\usepackage[T1]{fontenc}
\usepackage{wrapfig}
\usepackage[a4paper,top=3cm,bottom=2cm,left=3cm,right=3cm,marginparwidth=1.75cm]{geometry}

\usepackage{subfigure}
\usepackage{comment}
\usepackage{amsmath}
\usepackage{amsfonts}
\usepackage{multirow}
\usepackage{graphicx}
\usepackage[colorinlistoftodos]{todonotes}
\usepackage{authblk}
\usepackage{enumitem}
\title{Gerrymandering and fair districting in parallel voting systems}

\author[1]{Igor Mandric}
\author[2]{Igor Ro\c{s}ca}
\author[3]{Radu Buzatu}

\affil[1]{Department of Computer Science, University of California Los Angeles, Los Angeles, CA, USA}
\affil[2]{Institute of Ecology and Geography, Chi\c{s}in\v{a}u, Moldova}
\affil[3]{Department of Mathematics, Moldova State University, Chi\c{s}in\v{a}u, Moldova}

\date{\vspace{-5ex}}

\begin{document}
\maketitle
\thispagestyle{empty}

\vspace{1cm}

\begin{abstract}
    Switching from one electoral system to another one is frequently criticized by the opposition and is viewed as a means for the ruling party to stay in power. In particular, when the new electoral system is a parallel voting (or a single-member district) system, the ruling party is usually suspected of a biased way of partitioning the state into electoral districts such that based on a priori knowledge it has more chances to win in a maximum possible number of districts. In this paper, we propose a new methodology for deciding whether a particular party benefits from a given districting map under a parallel voting system. As a part of our methodology, we formulate and solve several gerrymandering problems. We showcased the application of our approach to the Moldovan parliamentary elections of 2019. Our results suggest that contrary to the arguments of previous studies, there is no clear evidence to consider that the districting map used in those elections was unfair.
\end{abstract}

\vspace{1cm}

\section{Introduction}

% Contributions:
%     - We propose a novel methodology for detecting whether a particular party, block or coalition of parties benefit from the political districting
%     - We formulate a mathematical model which can find an optimal political districting for a particular party (block, coalition)
%     - We apply our model to the case of the Republic of Moldova (elections 2019)

Political (re-)districting represents the task of partitioning a geographic area (e.g., a state or an administrative unit) into a given number of electoral districts subject to a predefined set of requirements \cite{niemi1978theory}. Frequently, the requirements are formulated with respect to demographic and/or geographic peculiarities of the area \cite{williams1995political}. For example, natural demographic requirements may refer to an approximately equal number of registered voters in each district. A common-sense geographical requirement is the contiguity and compactness of the districts \cite{williams1995political,garfinkel1970optimal}. In a democratic state, the main goal of political (re-)districting is creating {\em an unbiased} partitioning of the territory such that none of the political players benefit from it. If the opposite happens, then such a situation is commonly referred to as (partisan) {\em gerrymandering}.

The problem of political (re-)districting arose at the beginning of the nineteenth century in Massachusetts elections when a salamander-shaped district was proposed \cite{lewyn1993limit}. Such an ``unnatural'' district shape was evidence that the legislature attempted to influence the electoral outcome by manipulating the districts' borders. The term ``gerrymandering'' originates from the name of the famous Massachusetts governor at that time, Elbridge Gerry. Since then, there were multiple cases in history when gerrymandering was detected in the elections across many countries in the world \cite{backstrom1977issues,mcgann2016gerrymandering,malesky2009gerrymandering,halas2017functionality}. A lot of effort has been made by the researchers to develop methods for the detection, quantification, and prevention of gerrymandering \cite{lewyn1993limit,plener2018quantifying, veomett2018efficiency,tapp2019measuring}. One of the standard ways for the prevention of gerrymandering is avoiding the creation of salamander or eel-like shapes by optimizing the compactness of the districts. For example, one of the earliest methods formulated in the seminal paper of Hess in the 1960s used a computational model that sought to optimize the moment of inertia of the districts \cite{hess1965nonpartisan}. However, at that time it was not possible to use it in practice because there were no computers able to handle it. Since then, several computational models for solving the political (re-)districting problem have been proposed \cite{kalcsics2015districting, ricca2013political}. Due to the fact that this problem in many of its formulations is NP-hard \cite{altman1997automation}, not only exact \cite{nygreen1988european,nemoto2003modelling,li2007quadratic} but also a lot of heuristic-based methods \cite{hojati1996optimal,bozkaya2003tabu,ricca2008local} for tackling it were developed.

In this paper, we propose a novel methodology for detecting gerrymandering and determining a fair (re-)districting in parallel voting systems. As a part of our methodology, we formulate and solve two mathematical problems. The first problem consists of finding a political (re-)districting that maximizes (minimizes) the number of single-member districts won by a particular party and the second problem is finding a (re-)districting map with a specified electoral outcome. Finally, we applied our model to the case of the parliamentary elections held in the Republic of Moldova under the parallel voting system in 2019 to identify the political force with the highest benefit from the actual districting and drawing an example of a fair single-member district map.

\section{Background and motivation}

Parliamentary (assembly) elections are a process of selecting representatives in the main legislative body of government in a state. The two most frequently used electoral systems are {\em party-list proportional representation} and {\em parallel voting} ({\em mixed}) systems. In the family of party-list proportional representation systems, parties make lists of candidates, and the seats get distributed to each party proportionally to the number of votes received in the elections. Such systems are used, for example, in Argentina, Brazil, Finland, Poland, and Spain. Parallel voting systems differ from party-list proportional representation systems in that a part of the seats in a parliament are filled as the result of a single-member district election. Such systems are used in Russia, Hungary, Japan, and South Korea.

Interestingly, switching from a party-list proportional representation system to a parallel voting system can be viewed as a means of obtaining more seats in the parliament by the ruling party in the next elections (especially when its rating is constantly decreasing). One aspect of such accusations is that the ruling party may influence the process of drawing the districting map in a way that favors its electoral outcome. An example of such switching happened in Moldova in the 2019 parliamentary elections. The party at power (Democratic Party of Moldova) was accused by the opposition and the civil society of the non-transparent process of drawing the districting map. The democrats were suspected of gerrymandering, and several studies claimed that they detected gerrymandering in those elections (for example, \cite{pasha2019gerrymandering}). The main argument in favor of that claim was based on the fact that the Democrats received more seats in the single-member district elections than the number of seats they received in the nation-wide constituency with the proportional system. Such an approach for detecting gerrymandering is wrong since it does not take into account the geographical distribution of the voters' preferences (i.e., the demography of the country). Likewise, without taking into account this distribution it is not possible to construct a fair districting.

Detecting gerrymandering in an unbiased and demographically-aware way requires the comparison of the actual districting used in the elections with all the other possible ways to partition the state into single-member districts. Generating all possible districting maps is a computationally unfeasible task even for such a small country as the Republic of Moldova. Therefore, computational methods allowing to efficiently dissect the search space of all possible districting maps are critical for understanding whether a particular districting map is fair or favoring a selected political party.

\section{Methods}

\subsection{Projecting the results of the nation-wide constituency onto the actual districting}

Consider a country with a party-list proportional representation electoral system that is going to transition to a parallel voting system. 
Given a set of rules $\mathcal{R}$ specifying how the territory of the state is supposed to be partitioned into electoral districts, one may draw the electoral districts in multiple ways. Let $\mathcal{P}(\mathcal{R})$ be the set of all such electoral partitions. Different partitions $p_1, p_2 \in \mathcal{P}(\mathcal{R})$ may not be equivalent for a particular party. For example, based on a priori voting preferences of the population in different regions of the country (for example, based on the results of the previous elections), partition $p_1$ may be more favorable for this party than partition $p_2$. 

We postulate that a priori each citizen votes for the same party both in the nation-wide constituency and in the local district (we refer to this as the {\em voting postulate}). This means that without knowing the candidates representing each party, none of the voters has the interest to vote for a candidate of another party. Note that in practice, when the candidates are known, some citizens may switch their voting preference in the local districts in favor of a more notorious candidate representing another party.

Given the actual districting $p_{actual} \in \mathcal{P}(\mathcal{R})$, one can use the voting postulate to estimate the results of the single-district elections by projecting the results of the nation-wide constituency onto the partition $p_{actual}$. To detect gerrymandering, one has to enumerate all possible partitions in $\mathcal{P}(\mathcal{R})$ to determine the range of the possible number of seats each party can win in the single-district elections. If a party wins under $p_{actual}$ a number of seats close to its maximum value across all the partitions, then this might be an indication of gerrymandering.

\subsection{Detecting gerrymandering and drawing a fair districting map}

A theoretically attractive approach consisting of enumerating all the partitions in $\mathcal{P}(\mathcal{R})$ is not feasible in practice. Various attempts have been made to develop simulation methods allowing the generation of all possible districting maps \cite{fifield2015new, chen2016evaluating, tam2016toward, borodin2018big}. Since generating all districting maps is extremely challenging, all efforts in this direction have been successful only for relatively small states. Such models didn't offer the possibility to generate sufficiently informative and random partitions at the state level within a reasonable time. For example, the well-known 538 Gerrymandering Project \cite{538} uses 2568 districting maps drawn by hand to detect gerrymandering in the United States. This methodology is tedious and unreliable because a small sample of districting maps can not provide an adequate snapshot of all possible districting partitions. Furthermore, gerrymandering is not a probabilistic process and if a party influenced the process of drawing the districting map it would rather use an optimization technique to determine the most favorable one for itself instead of randomly sampling from $\mathcal{P}(\mathcal{R})$.

%makes the problem prodigiously larger and unrealistic on their computing platform.

%Enumerating all the partitions in $\mathcal{P}(\mathcal{R})$ is not feasible. For example, the well-known 538 Gerrymandering Project \cite{538} uses 2568 districting maps drawn by hand in order to detect gerrymandering in the United States.

%We consider that two different partitions $p_1, p_2 \in \mathcal{P} (\mathcal{R})$ are equivalent if under the voting postulate they yield the same {\em electoral outcome} in the single-member district elections. We define an electoral outcome as a tuple $v = (v_0, v_1, ..., v_m, ..., v_{M - 1})$ where $v_m$ it the number of seats won by party $m$ and the sum of all $v_m$ across the $M$ parties is equal to the number of single-member districts $K$. We denote by $V$ the set of all possible outcomes. Let $\mathcal{V}_{m_0}$ be the sorted list of the number of seats party $m_0$ wins in the single-member district elections under all possible outcomes in $V$. Given the electoral outcome $v_{actual}$ (corresponding to the partition $p_{actual}$), we define the {\em efficiency score} $E_{m_0}$ of party $m_0$ as the average percentile score of $v^{actual}_0$ in $\mathcal{V}_{m_0}$. If party $m_0$ has the highest efficiency score under $p_{actual}$ (among the other parties) then we consider that $p_{actual}$ is the most favorable for $m_0$.

We consider that two different partitions $p_1, p_2 \in \mathcal{P} (\mathcal{R})$ are equivalent if under the voting postulate they yield the same {\em electoral outcome} in the single-member district elections. We define an electoral outcome as a tuple $o = (o_1, ..., o_m, ..., o_M)$ where $o_m$ is the number of seats won by party $m$ and the sum of all $o_m$ across the $M$ parties is equal to the number of single-member districts $K$. We denote by $\mathcal{O}^{*}$ the set of all possible outcomes and by  $\mathcal{O}$ the set of all feasible outcomes. Note that for an outcome $o$ to be feasible, at least one partition with outcome $o$ must exist in $\mathcal{P}(\mathcal{R})$. 

%Let $\mathcal{O}_{m_0}$ be a list of natural numbers that is obtained in the following way: 
%\begin{enumerate}
%    \item Initially, $\mathcal{O}_{m_0}$ is empty;
%    \item We add the number of seats $o_m$ to $\mathcal{O}_{m_0}$ for each %electoral outcome $o\in \mathcal{O}$;
%    \item Sort the list $\mathcal{O}_{m_0}$ such that the smallest value %has the index 1 and the largest value has the index $|\mathcal{O}_{m_0}|$.
%\end{enumerate}

%In other words, 

Let $\mathcal{O}_{m}$ be the sorted list of the number of seats party $m$ wins in the single-member district elections under all feasible outcomes in $\mathcal{O}$. We define the {\em efficiency score} $e_{m}(o)$ of party $m$ in electoral outcome $o\in \mathcal{O}$ as the average percentile score of $o_m$ in $\mathcal{O}_{m}$:

\begin{equation}
    e_{m}(o)= \sum_{k\in L_{\mathcal{O}_{m}}(o_m)} \frac{k-0.5}{N\times l} \times 100\%
\end{equation}
%*
where $L_{\mathcal{O}_{m}}(o_m)$ is the set of all indexes of $o_m$ in list $\mathcal{O}_{m}$,  $l=|L_{\mathcal{O}_{m}}(o_m)|$, and $N$ is the overall size of $\mathcal{O}_{m}$. The intuitive interpretation of the efficiency score is the following: if under some partition $p \in \mathcal{P}(\mathcal{R})$ party $m$ wins its minimal number of districts then its efficiency score is 0, and if it wins its maximal number of districts then its efficiency score is 100.

Enumerating the elements of $\mathcal{O}$ is a less challenging problem than enumerating all the partitions in $\mathcal{P}(\mathcal{R})$. To do so, we need to find the minimum $o^{min}_m$ and the maximum $o^{max}_m$ number of seats for each party $m$ and then find all the tuples $o$ for which $o^{min}_m \leq o_m \leq o^{max}_m, \; m = 1, 2, ..., M$ and $\sum\limits_{m = 1}^M o_m = K$ and for which at least one partition in $\mathcal{P}(\mathcal{R})$ exists. However, computing $o^{min}_m$ and $o^{max}_m$ is not a  trivial task. We refer to the problem of computing $o^{max}_m$ ($o^{min}_m$) as the Maximum (Minimum) Gerrymandering Problem. Proving the feasibility of an outcome $o \in \mathcal{O}$ is also a challenging task. We refer to this problem as the Fixed Gerrymandering Problem. In Section \ref{ilp} we formulate and solve these problems.

We define an {\em absolutely fair partition} as a partition for which the efficiency score of each party is equal to 50. An absolutely fair partition provides no advantage to any party in the single-member district elections. An absolutely fair partition might not necessarily be feasible. A {\em fair partition}, denoted $p_{fair}$, is such a partition for which its efficiency profile is closest to the efficiency profile of an absolutely fair partition (in $L_1$ sense), i.e.  $p_{fair}$ is a partition for which the electoral outcome $o_{fair}$ is equal to

\[
\operatorname*{argmin}_{o\in\mathcal{O}} \sum_{m=1}^{M} |e_{m}(o) - 50|
\]

%is minimal on the set of all feasible partitions, where $o_{fair}$ is the electoral outcome of $p_{fair}$.

Thus, deciding whether a particular partition $p \in \mathcal{P}(\mathcal{R})$ is fair or biased towards a particular party can be done with the following procedure:

\begin{enumerate}
    \item For each party $m$, solve the Minimum and the Maximum Gerrymandering Problems to identify the minimum $o^{min}_m$ and the maximum $o^{max}_m$ possible number of seats won by $m$ in the single-member district elections.
    \item Identify the set of all potentially possible outcomes $\mathcal{O}^{*}$.
    \item For each $o^{*} \in \mathcal{O}^{*}$, solve Fixed Gerrymandering Problem to find out whether $o^{*}$ is feasible. Select only the set of feasible outcomes $\mathcal{O}$.
    
    \item For each outcome $o\in \mathcal{O}$, compute the efficiency score of each party $m$ and the $L_1$ distance between the efficiency profile corresponding to $o$ and the efficiency profile of the outcome corresponding to an absolutely fair partition. 
    
    \item Sort the set of feasible outcomes $\mathcal{O}$ by the distances computed in step 4. If the outcome corresponding to $p$ is in the top 5\% of outcomes closest to the absolutely fair one then there is no evidence of gerrymandering and partition $p$ yields a relatively fair districting map. Otherwise, the party with the highest efficiency score can be suspected of gerrymandering.

    %\item Sort the set of outcomes $\mathcal{O}$ by distances from the previous stage, and determine the divergences of each partition from an absolutly  by computing proportion of  divergence from of each ouctcome the proportion of divergence.
    
    %\item For each party $m$, compute its efficiency score assuming the outcome corresponding to $p_{actual}$. Find the $L_1$ distance between the efficiency profile of $p_{actual}$ and the efficiency profile of an absolutely fair partition.
\end{enumerate}

\section{Gerrymandering problems}\label{ilp}

\subsection{Minimum and Maximum Gerrymandering Problems}

In a single-member district system, party $m$ wins an electoral district if it accumulates more votes than any other participating party. As a result of winning elections in a district, party $m$ receives a seat in the parliament. The goal of each party is to maximize its number of seats. Having full control over the process of partitioning the state into a set of electoral districts, party $m$ would adopt a partition $p_{m} \in \mathcal{P}(\mathcal{R})$ which would fully advantage its electoral outcome. Therefore, it would solve the following:

\vspace{0.2cm}
{\noindent\bf Maximum Gerrymandering Problem}. The state consists of $N$ localities. Each locality $i, \; i = \overline{1, N}$ is characterized by the total number of registered voters $p_i$. $M$ political parties participate in the elections with $K$ electoral districts under a parallel voting system. Given $v_i^m$, the number of votes party $m, \; m = \overline{1, M}$ wins in the locality $i, \; i = \overline{1, N}$, determine a partition $p_m \in \mathcal{P}(\mathcal{R})$ of the state into $K$ electoral districts such that party $m$ wins the maximum possible number of districts. The set of rules $\mathcal{R}$ is:

\begin{itemize}
    \item The centers of the $K$ districts are predefined (the center of the district $k$ is denoted as $C_k, \; k = \overline{1, K}$). These centers are, for example, the largest localities in the country;
    \item The $K$ districts must be geographically contiguous, i.e, there should exist a path between any two localities in the district, two localities being neighbors if and only if they share a border;
    \item The number of registered voters in each district should not be less than $A$ and not more than $B$.
\end{itemize}
\vspace{0.2cm}

Solutions of the Maximum Gerrymandering Problem (it is possible to have multiple solutions with the same value of the objective function) yield the upper bound on the number of won districts due to gerrymandering. To obtain the lower bound, one has to solve the Minimum Gerrymandering Problem. We will refer to these problems as Max-GP (similarly, Min-GP).

\subsection{Fixed Gerrymandering Problem}

Proving that a particular electoral outcome $o \in \mathcal{O}^{*}$ is feasible can be done by giving an example of a partition $p \in \mathcal{P}(\mathcal{R})$ yielding $o$. In other words, one has to solve the following

\vspace{0.2cm}

{\noindent\bf Fixed Gerrymandering Problem}. Given the setup of Max-GP (or Min-GP), determine if there exists at least one partition $p \in \mathcal{P}(\mathcal{R})$ yielding $o$.  

\vspace{0.2cm}

In the next section, we propose an integer linear programming (ILP) approach for solving min-GP (max-GP) and fixed-GP.

\subsection{Integer Linear Programming approach}

We view the localities of the state as the nodes $V$ of an undirected graph $G = (V, E)$. Two nodes $i$ and $j$ are connected by an edge $e_{ij} \in E$ if and only if the two localities $i$ and $j$ are geographically adjacent, i.e. share a border. We will refer to $G$ as {\em geographical adjacency graph}. We introduce binary variables $u_{ik}$ with the following meaning:
\begin{equation}
  u_{ik}=\begin{cases}
    1, & \text{if locality $i$ is assigned to district $k$} \\
    0, & \text{otherwise}.
  \end{cases}
\end{equation}

Each locality is assigned to exactly one district. This can be described by the following equation:

\begin{equation}
    \sum\limits^K_{k=1}u_{ik} = 1, \; i = \overline{1, N}
\end{equation}

%\begin{equation}
%    \sum\limits^N_{i=1}u_{ik} = 1, \; k = \overline{1, K}
%\end{equation}

To guarantee that each district is contiguous, i.e. subgraph $G_k$ induced by the localities assigned to district $k$ is connected, we use the following flow formulations. For each edge $e_{ij} \in E$ and for each district $k$, we introduce integer-valued non-negative flow variables $F_{ijk} ,\; i, j = \overline{1, N}, \; k = \overline{1, K}$. We also introduce two additional vertices $s$ and $t$ to the graph $G$ - the source and the sink. The variables $F_{ijk}$ must satisfy multiple conditions. First of all, they should be bounded by a large enough number $L$:

\begin{equation}
    F_{ijk} \leq L, \; e_{ij} \in E, \; i, j = \overline{1, N}, \; k = \overline{1, K}.
\end{equation}

The flow from source $s$ to the nodes of subgraph $G_k$ propagates from $s$ only to the center $C_k$ of district $k$ and it is equal to the number of localities assigned to district $k$. This can be described by the following two conditions:

\begin{equation}
    F_{sik} = 0, \; k = \overline{1, K}, \; i = \overline{1, N}, \; i \neq C_k
\end{equation}

and

\begin{equation}
    \sum\limits_{i = 1}^{N} F_{sik} = \sum\limits_{i = 1}^{N} u_{ik}, \; k = \overline{1, K}
\end{equation}

The flow from sink $t$ to all the nodes of graph $G$ is 0:

\begin{equation}
    F_{tik} = 0, \; i = \overline{1, N}, \; k = \overline{1, K}
\end{equation}

For each subgraph $G_k$, the total flow which goes to sink $t$ is also equal to the number of localities assigned to district $k$:

\begin{equation}
    \sum\limits_{i = 1}^{N} F_{itk} = \sum\limits_{i = 1}^{N} u_{ik}, \; k = \overline{1, K}
\end{equation}

At each node, the conservation of flow must be satisfied:

\begin{equation}
    F_{sik} + \sum\limits_{j = 1}^{N} F_{jik} = F_{itk} + \sum\limits_{j = 1}^{N} F_{ijk}, \; i = \overline{1, N}, \; k = \overline{1, K}
\end{equation}

Next, if an edge $e_{ij} \in E$ connects two localities assigned to different districts then the flow $F_{ijk}$ through this edge in district $k$ is null. This is described by the following two conditions:

\begin{equation}
    F_{ijk} \leq L \cdot u_{ik}, \; e_{ij} \in E, \; i, j = \overline{1, N}, \; k = \overline{1, K}
\end{equation}

and 

\begin{equation}
    F_{ijk} \leq L \cdot u_{jk}, \; e_{ij} \in E, \; i, j = \overline{1, N}, \; k = \overline{1, K}
\end{equation}

The requirement that the total number of registered voters of district $i$ is ranged between $A$ and $B$ is described by the following restrictions:

\begin{equation}
    \sum\limits_{i = 1}^{N} d_i u_{ik} \geq A, \; k = \overline{1, K}
\end{equation}

and

\begin{equation}
    \sum\limits_{i = 1}^{N} d_i u_{ik} \leq B, \; k = \overline{1, K}
\end{equation}

where $d_i$ is the number of registered voters in locality $i$.

Party $m$ dominates another party $m'$ in district $k$ if $m$ accumulates a strictly greater number of votes than $m'$ in district $k$. We introduce binary variables $w^m_{m'k}$ with the following meaning:

%Party $m_0$ dominates another party $m$ in district $k$ if $m_0$ accumulates a strictly greater number of votes than $m$ in district $k$. We introduce binary variables $w_{mk}$ with the following meaning:

\begin{equation}
  w^m_{m'k}=\begin{cases}
    1, & \text{if party $m$ dominates party $m'$ in district $k$} \\
    0, & \text{otherwise}.
  \end{cases}
\end{equation}

%\begin{equation}
%  w_{mk}=\begin{cases}
%    1, & \text{if party $m_0$ dominates party $m$ in district %$k$} \\
%    0, & \text{otherwise}.
%  \end{cases}
%\end{equation}

The domination of party $m$ over party $m'$ (i.e., when party $m$ accumulates strictly more votes than party $m'$) in district $k$ can be described by the following conditions:

%The domination of party $m_0$ over party $m$ in district $k$ can be described by the following conditions ($L$ is a large enough number):

\begin{equation}
    \sum\limits_{i = 1}^{N} v^m_i u_{ik} \; \geq \; \sum\limits_{i = 1}^{N} v^{m'}_i u_{ik} + 1 - L \cdot (1 - w^m_{m'k}),
\end{equation}\label{eq1-bigger}

%\begin{equation}
%    \sum\limits_{i = 1}^{N} v^{m_0}_i u_{ik} \; \geq \; %(\sum\limits_{i = 1}^{N} v^m_i u_{ik} + 1) + 1 - L \cdot (1 - %w_{mk}),
%\end{equation}\label{eq1-bigger}

and

\begin{equation}
    \sum\limits_{i = 1}^{N} v^{m'}_i u_{ik} \; \geq \; \sum\limits_{i = 1}^{N} v^m_i u_{ik} - L \cdot w^m_{m'k}, 
\end{equation}\label{eq2-bigger}

%\begin{equation}
%    (\sum\limits_{i = 1}^{N} v^m_i u_{ij} + 1) \; \geq \; %\sum\limits_{i = 1}^{N} v^{m_0}_i u_{ik} + 1 - L \cdot w_{mk}, 
%\end{equation}\label{eq2-bigger}

where $k = \overline{1, K}, \; m, m' = \overline{1, M}, \; m \neq m'$.

%where $k = \overline{1, K}, \; m = \overline{1, M}, \; m \neq m_0$.

%Party $m_0$ wins in district $k$ if and only if it dominates all other parties $m, \; m = \overline{1, M}, \; m \neq m_0$ in district $k$. This can be expressed by the following two conditions:

Party $m$ wins in district $k$ if and only if it dominates all other parties $m', \; m' = \overline{1, M}, \; m' \neq m$ in district $k$. This can be expressed by the following two conditions:

\begin{equation}
    w^m_{m'k} \geq w^m_k, \; m, m' = \overline{1, M}, \; m \neq m',  \; k = \overline{1, K}
\end{equation}

%\begin{equation}
%    w_{mk} \geq w_k, \; m = \overline{1, M}, \; k = %\overline{1, K}
%\end{equation}

and

%\begin{equation}
%    \sum\limits_{\substack{m = 1 \\ m \neq m_0}}^{M} w_{mk} %\leq w_k + K - 2,  \; k = \overline{1, K}
%\end{equation}

\begin{equation}
    \sum\limits_{\substack{m' = 1 \\ m' \neq m}}^{M} w^m_{m'k} \leq w^m_k + M - 2, \; m = \overline{1, M}, \; k = \overline{1, K}
\end{equation}

where 

\begin{equation}
  w^m_{k}=\begin{cases}
    1, & \text{if party $m$ wins in district $k$} \\
    0, & \text{otherwise}.
  \end{cases}
\end{equation}

%\begin{equation}
%  w_{k}=\begin{cases}
%    1, & \text{if party $m_0$ wins in district $k$} \\
%    0, & \text{otherwise}.
%  \end{cases}
%\end{equation}

The objective function for max-GP and min-GP is:

\begin{equation}
    \sum\limits_{k = 1}^{K} w^m_{k} \rightarrow max \; (min) \\
\end{equation}

For solving fixed-GP, we introduce additional constraints:

\begin{equation}
    \sum\limits_{k = 1}^K w^m_k = o_m,  \; m = \overline{1, M}
\end{equation}

and the objective function is a constant (since we are searching for a feasible solution).

\section{Case study: Moldova parliamentary elections 2019}

In this section, we applied our method to determine the party for which the actual partition of the state into single-member districts was most favorable in Moldovan parliamentary elections which took place in February 2019 under the parallel voting system. We also found a fair districting and determine whether the actual districting can be considered fair.

\subsection{The electoral system of Moldova}

Before 2019, the 101 members of Parliament were elected by party-list proportional representation system. After a reform of the electoral system, the elections in 2019 were held under the parallel voting system. Each citizen was supposed to vote twice - once for a party in the nation-wide constituency and once for a candidate (either a representative of a party or an independent candidate) in his/her local district. The elections in the nation-wide constituency accounted for 50 seats in the Parliament and the single-member district elections accounted for the other 51 seats. In each district, the candidate with the largest number of votes was considered to be the winner.

%As Moldova consists of 34 territorial districts (which is less than 51), A special committee for determining the partition of Moldova into 51 electoral districts was was convoked. 

%A special committee for determining the partition of Moldova into 51 electoral districts was convoked. As a result of the work of the committee, 9 districts were allocated for for the capital of Moldova, Chisinau and 2 for its suburbs, 3 districts were allocated for so-called ``diaspora'' (USA and Canada, Europe, Russian Federation), 2 districts were allocated for Gagauzia, 1 district for the compact Bulgarian community in Taraclia, 2 districts for Transnistria (a separatist region), and 32 districts were allocated to the rest of Moldova. Much controversy was spurred after the actual electoral district partition was made publicly available. In particular, then-at-power Democrat party (in collaboration with then-at-opposition Socialist party) was accused to influence the process of electoral partitioning so that to a priori boost their chances to win in the maximum number of local districts. 

%According to the electoral code of the Republic of Moldova, parliamentary elections since 2019 are conducted based on a parallel voting system  (proportional and majoritarian) in one national constituency and within Single Member Constituencies. The 50 members of Parliament are elected in the national constituency on the basis of proportional representation, and the other 51 members of Parliament are elected within Single Me Constituencies by majority vote, one member from each constituency.

As the result of the reform, the following territorial and demographic criteria for delineating electoral districts were set out in the Electoral Code of Moldova:

\begin{enumerate}[label=(\alph*)]
    \item each district must have at least 55,000 and at most 60,000 voters;
    \item deviation of the number of voters among districts is not allowed to exceed 10\%;
    \item if the number of voters of a locality is higher than the average number calculated for an electoral district then several districts have to be established on the basis of this locality. Otherwise, setting boundaries of electoral districts within a locality is prohibited;
    \item districts where national minorities live densely are established, taking into account the boundaries of the respective localities;
    \item districts on the territory of Autonomous Territorial Unit of G\v{a}g\v{a}uzia are established within its boundaries.
  \end{enumerate}

{\noindent\bf The actual districting in 2019 elections.} In the elections, 3 districts were associated with the Moldovan citizens living abroad (the so-called ``diaspora''), 2 districts were allocated to the Transnistria region, and the rest 46 out of 51 electoral districts were located on the territory controlled by the Moldovan government. Moldova was conventionally divided into 5 parts: 1) 9 districts allocated to the capital - Chi\c{s}in\v{a}u City, 2) 2 districts allocated to Chi\c{s}in\v{a}u suburbs, 3) 2 districts allocated to G\v{a}g\v{a}uzia, 4) 1 district allocated to the Bulgarian compact community of Taraclia, 5) the remaining 32 districts (we refer to them as $MD_{32}$). All the districts except the ones located in Chisinau were assigned an informal center. The map corresponding to the actual districting can be accessed on the official web-page of the Moldovan government \cite{harta}.

{\noindent\bf Political parties.} In 2019, 15 political forces (14 parties and 1 alliance) participated in the elections. We considered only four political forces: Democratic Party of Moldova (PDM), Party of Socialists of the Republic of Moldova (PSRM), political alliance ACUM consisting of two parties - Action and Solidarity Party (PAS) and Dignity and Truth Platform Party (PPDA), and \c{S}OR party. Other parties did not succeed in the Parliament.

\subsection{Results}

In our analysis, we used only publicly available data. We downloaded the results of the elections in the nation-wide constituency from the official web-page of the Central Electoral Commission of the Republic of Moldova \cite{cecmd}. 

%The geographical adjacency graph was built based on the map from the official web-page of the Moldovan government \cite{harta}. 

%In the elections, 3 districts were associated with the Moldovan citizens living abroad (the so-called ``diaspora''), 2 districts were allocated to the Transnistria region, and the rest of 46 out of 51 electoral districts were located on the territory controlled by the Moldovan government. Further in the paper, we will report our results based only on these 46 districts. Moldova was conventionally divided into 5 parts: 1) 9 districts allocated for the capital - Chi\c{s}in\v{a}u City, 2) 2 districts allocated for Chi\c{s}in\v{a}u suburbs, 3) 2 districts allocated for G\v{a}g\v{a}uzia, 4) 1 district allocated for the Bulgarian compact community of Taraclia, 5) the remaining 32 districts (we refer to them as $MD_{32}$). 

According to the nation-wide results, the overwhelming majority of votes in G\v{a}g\v{a}uzia were in favor of PSRM, and other parties under any possible districting maps have no success there. Likewise, Taraclia was won by PSRM with a large margin. In Chi\c{s}in\v{a}u suburbs, the overwhelming majority of votes belong to ACUM, and no districting map can give any advantage to other political parties. Thus, we concentrated on 2 regions: Chi\c{s}in\v{a}u City and $MD_{32}$.

\begin{figure}
    \centering
    \includegraphics[scale=0.45]{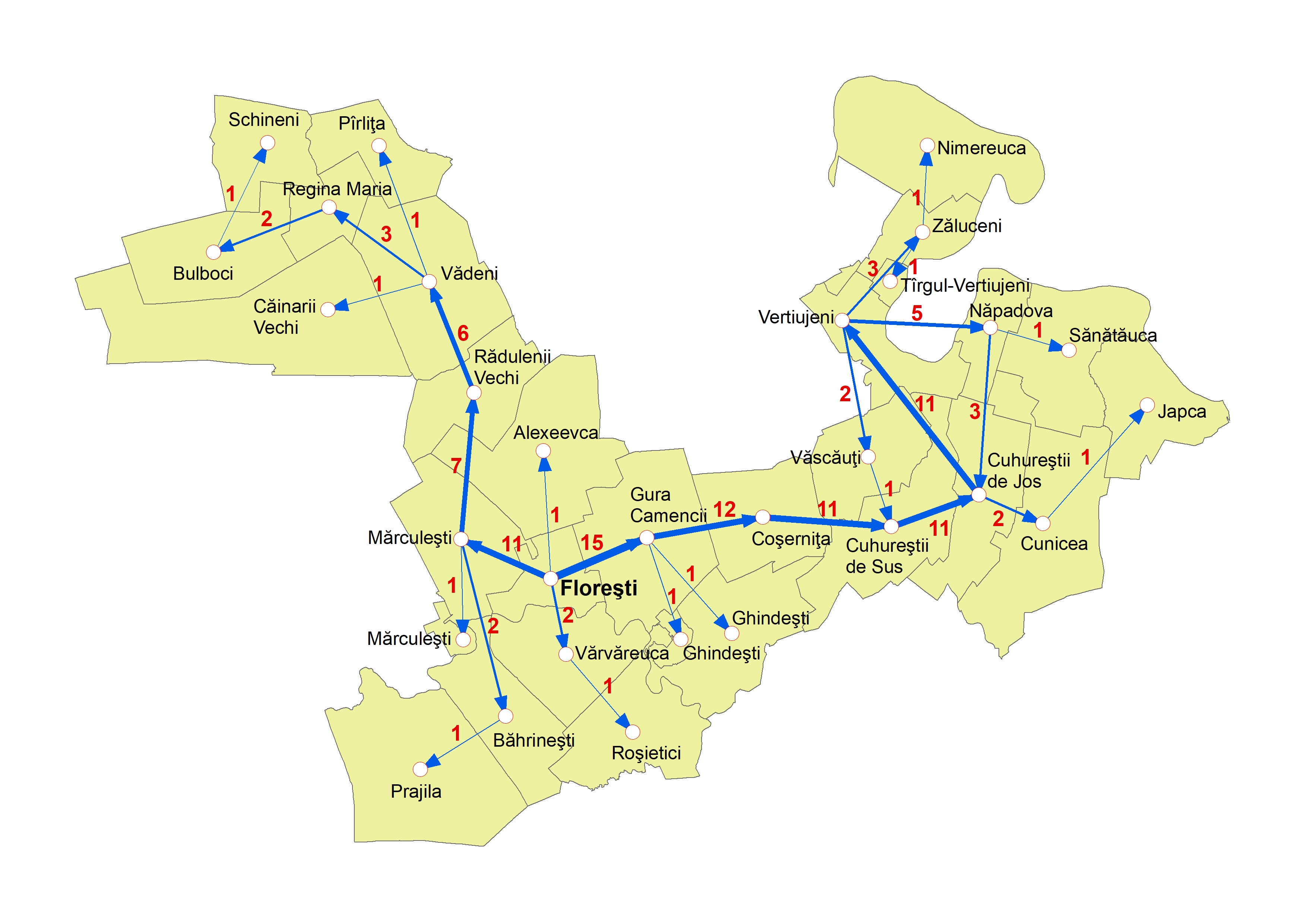}
    \caption{Flore\c{s}ti district and its flow (PSRM Maximum map). The flow is obtained as a part of the solution of the Max-GP problem and it guarantees that the district is contiguous.}
    \label{fig:florestiu}
\end{figure}

%We applied our model independently to Moldova region (32 districts) and Chisinau (9 districts) since the results of elections in these regions did not influence each other. We did not apply our model to Taraclia district (since it is won by PSRM), to Chisinau suburbs (since it is won by ACUM), and to Gagauzia (since it is won by PSRM). Owing to the lack of statistical and geographical data, we removed the territory of Transnistria and ``diaspora'' constituencies from our analysis.

We implemented the integer linear programs for min-GP, max-GP, and fixed-GP using Python programming language and CPLEX solver (version 12.7). ILP approaches are not scalable to large instances of the gerrymandering problems, thus we split $MD_{32}$ region (32 districts, 838 localities)  into 4 parts each having 8 centers: 1) North, 2)  South, 3) West, 4) East. The split was made using the borders of the actual electoral districting so that it is one of the feasible solutions. To further speedup the computations, for each district center, we kept only the edges with the localities whose distance is in top 3 distances from the center. To avoid geographically unreasonable solutions that are non-compact (i.e, with eel-like shapes), we used an additional constraint that the total lengths of the borders of the districts should not exceed the total lengths of the borders of the actual districts. All the districts in every solution are contiguous (Figure \ref{fig:florestiu}).

\begin{figure}
    \centering
    \includegraphics[scale = 0.5]{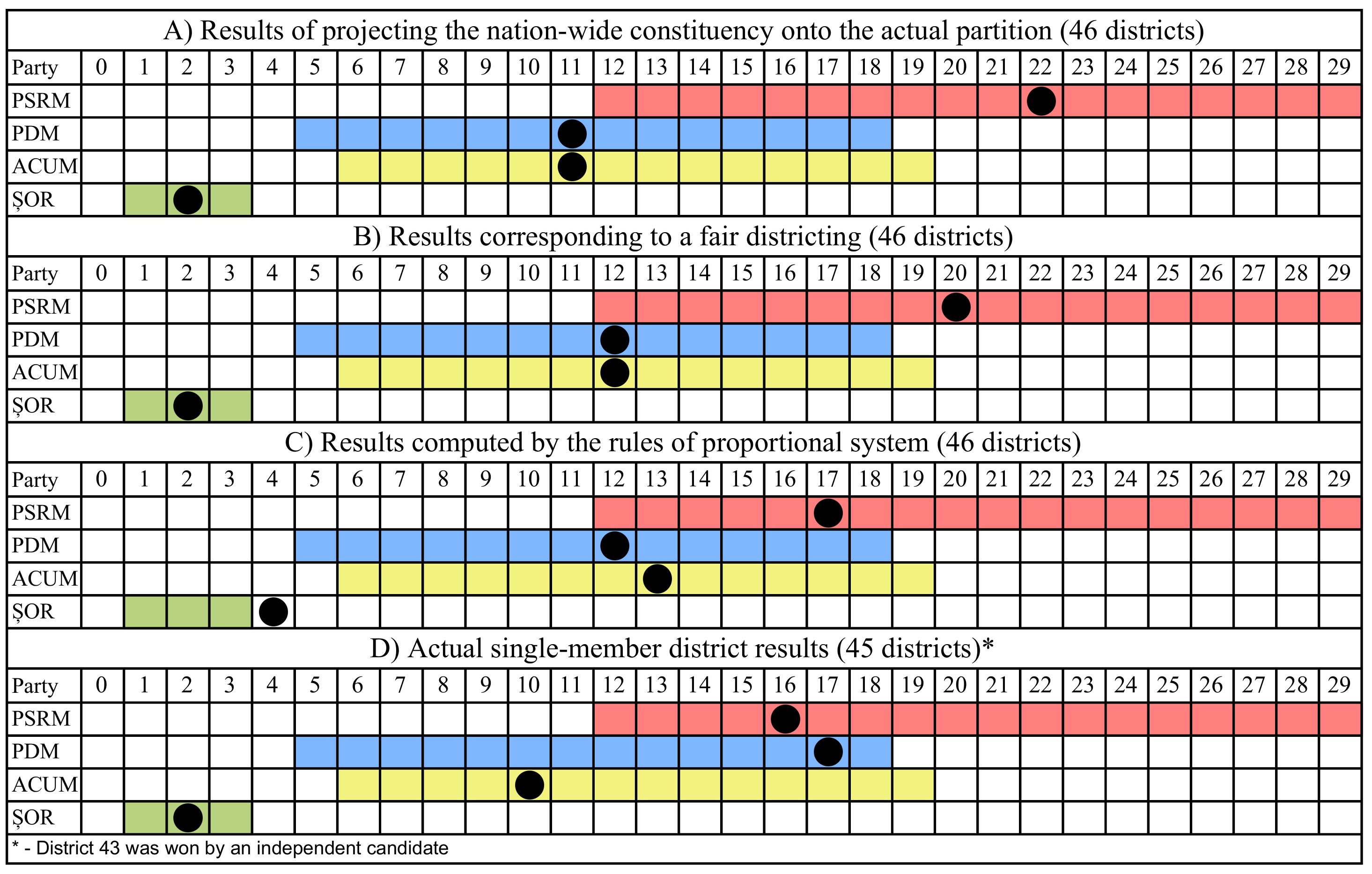}
    \caption{The electoral results (black dots) and the theoretical ranges of the number of won districts for each party.}
    \label{fig:fig2}
\end{figure}

For each party, we computed the theoretical bounds on the number of won districts in the single-district member elections. The lower bound (the minimal value obtained by solving Min-GP) corresponds to the most pessimistic districting scenarios, and there is no possibility to win a smaller number of districts. The upper bound (the maximum value obtained by solving Max-GP) corresponds to the most favorable scenarios, and better solutions do not exist. None of the four parties can lose all its votes due to the manipulation with the district's boundaries: PSRM wins at least 12 districts, PDM - 5, ACUM - 6, and \c{S}OR - 1 (see Figure \ref{fig:fig2}). Also, none of the parties can win all of the districts due to gerrymandering. In the best scenario for PSRM, it can potentially achieve victory in 29 out of 46 districts. Likewise, PDM, ACUM, and \c{S}OR in their most beneficial scenarios win 18, 19, and 3 districts correspondingly. Examples of the districtings for which each party wins its minimal and maximal number of districts are shown in Figures \ref{fig:max-acum}-\ref{fig:min-sor} (see also Supplementary Table 1).

\begin{figure}
    \centering
    \includegraphics[scale = 0.16]{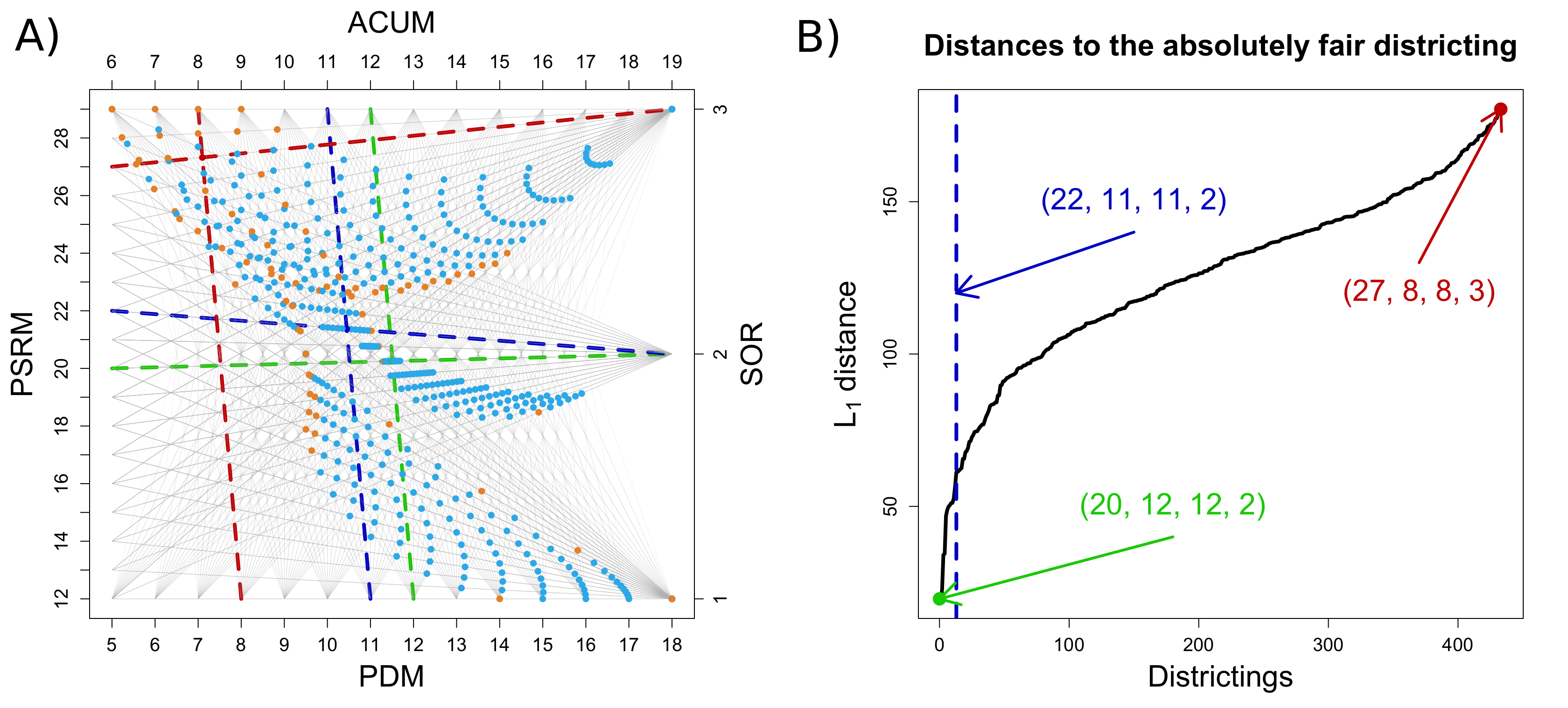}
    \caption{Visualization of all possible electoral outcomes. A) Each outcome is presented as an intersection of two lines. Light blue dots are feasible outcomes, and brown dots are infeasible outcomes. Green lines correspond to the fair outcome, red lines correspond to the outcome which is furthest from fair, blue lines correspond to the outcome projected onto the actual districting map. B) All the feasible outcomes sorted by the $L_1$ distance to the absolutely fair outcome. The three emphasized outcomes are the same as in A).}
    \label{fig:fig3}
\end{figure}

\begin{figure}[ht]
  \centering
  \includegraphics[scale=0.35]{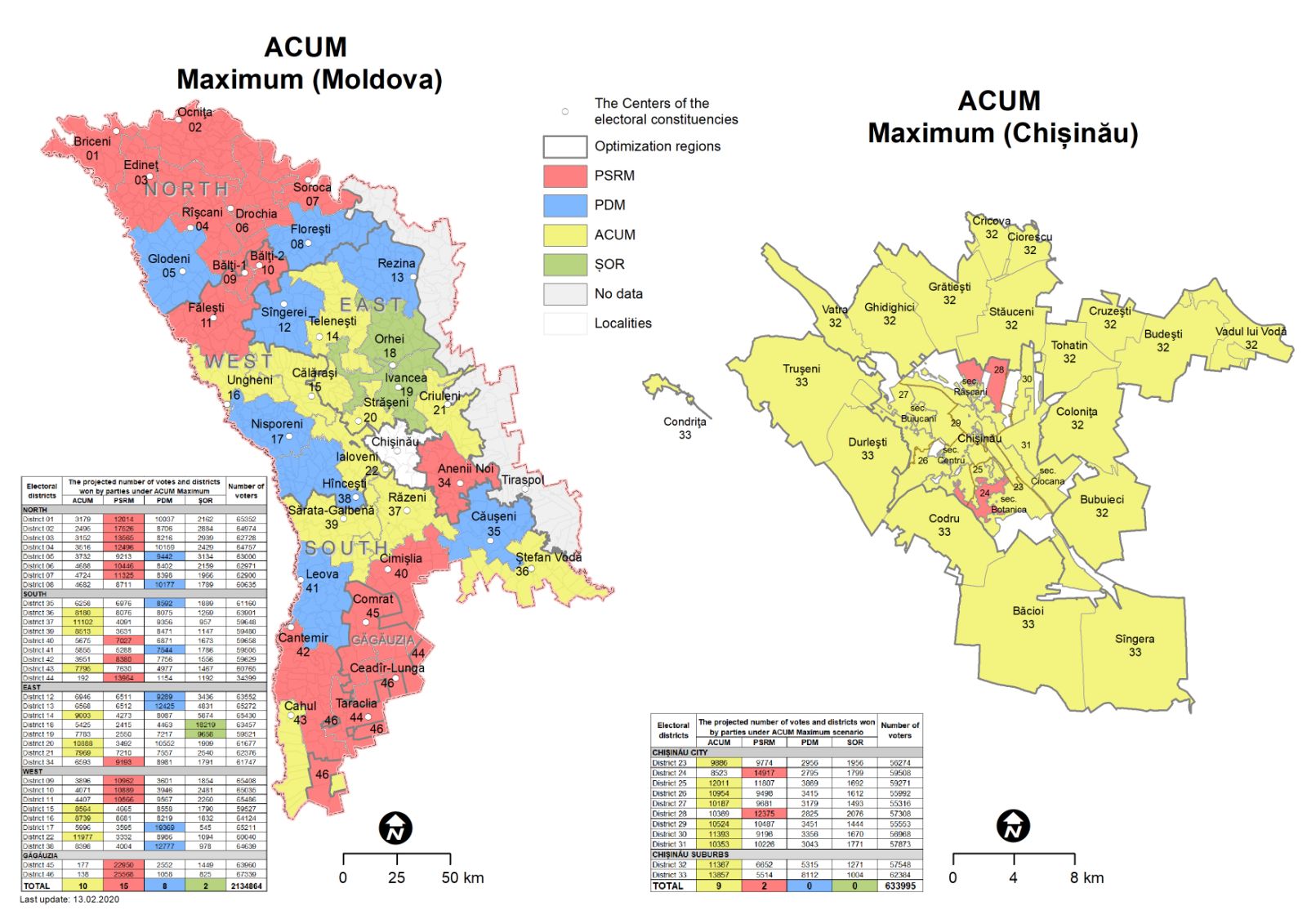}
  \caption{A maximum scenario for ACUM.}
  \label{fig:max-acum}

  \vspace*{\floatsep}

  \includegraphics[scale=0.35]{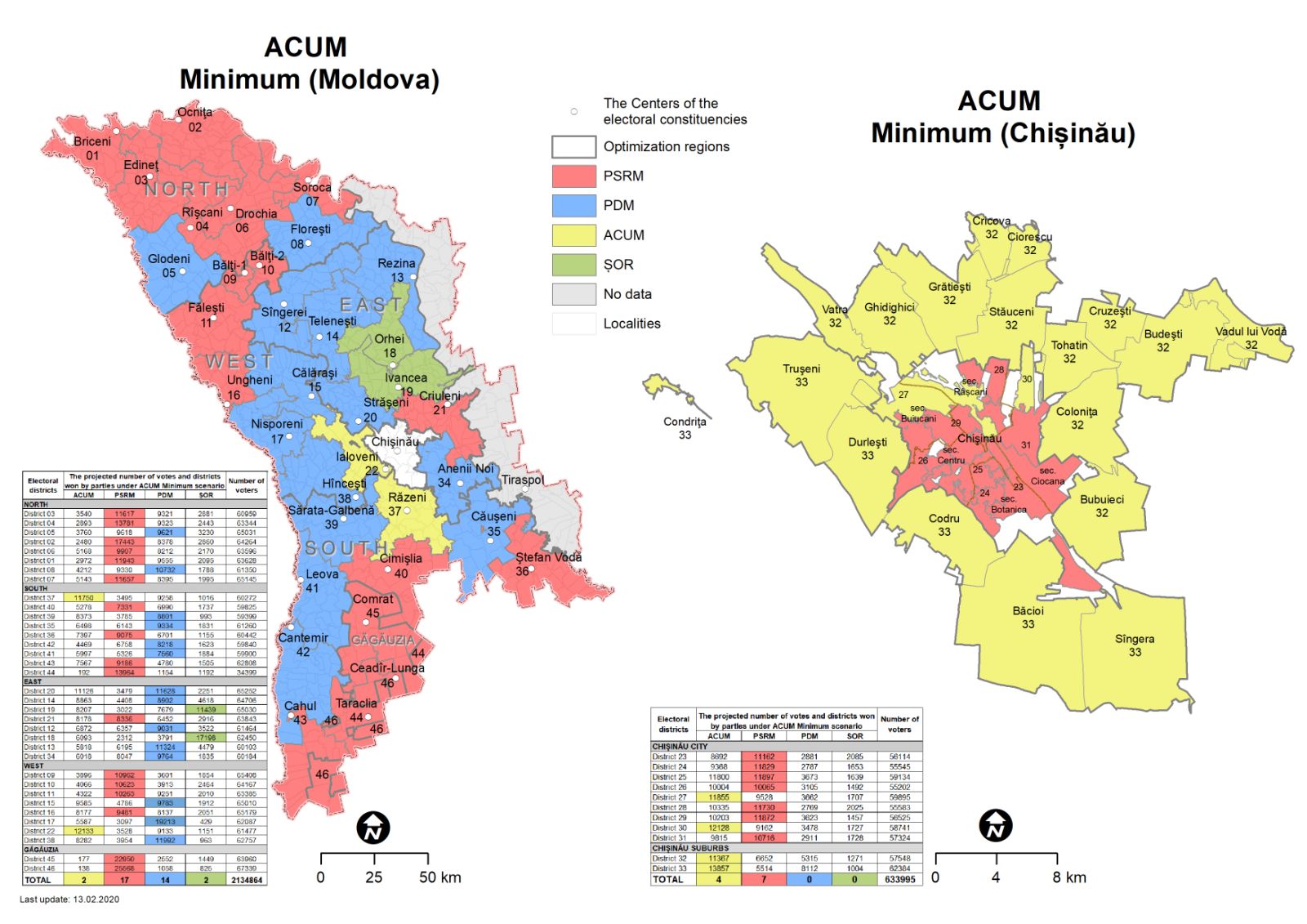}
  \caption{A minimum scenario for ACUM.}
  \label{fig:min-acum}
\end{figure}

Next, we identified all feasible electoral outcomes by solving fixed-GP. In total, out of 511 possible combinations of four numbers with ranges corresponding to each party's range and summing up to 46, only 433 are feasible (Figure \ref{fig:fig3}A). Note that 78 districtings are not feasible due to peculiarities of the distribution of voting preferences. For example, we found that in the East region, no districting admits both PSRM and \c{S}OR to win 3 districts (out of 8).

We projected the results of the nation-wide constituency onto the actual districting (Figure \ref{fig:actual}). In this case, PSRM wins 22 districts, PDM and ACUM each win 11 districts, and \c{S}OR wins 2 districts (Figure \ref{fig:fig2}A). Visually, these numbers fall close to the middle of each party's range of theoretically possible results. For an exact answer on who has the highest benefit from this districting, we computed the efficiency scores. PSRM has the highest efficiency score - 75.5\%, followed by \c{S}OR with 54.3\% and PDM with 40.7\%. The most disadvantaged electoral competitor is ACUM with only a 27.9\% efficiency score.

We next identified the outcome of a fair districting (Figure \ref{fig:fair}). Such a districting can be characterized by PSRM winning 20 districts, both PDM and ACUM - 12 districts each, and \c{S}OR - 2 districts (Figure \ref{fig:fig2}B). The efficiency scores, in this case, would be the following: PSRM - 49.3\%, PDM - 57.3\%, ACUM - 42.7\%, and \c{S}OR - 54.3\%. The $L_1$ distance of such an outcome to the outcome of the absolutely fair districting is 19.7, and all other outcomes are less fair (see Supplementary Table 2). 

Comparing the outcomes of a fair districting and the actual districting, one can observe that to make the actual districting fair, it is sufficient to change it in such a way that PSRM's outcome be less by 2 districts, but PDM's and ACUM's outcomes be more by 1 district. We found that such an update is possible in the South region. An example of a fair districting is provided in Figure \ref{fig:fair}. This districting is only different from the actual one by assigning 26 localities to different districts, or only around 3\% of the total number of localities (in terms of number of registered voters, it is only 1.6\%). To answer the question whether the actual districting can be considered fair, we sorted all the possible outcomes by the $L_1$ distance of their efficiency profiles to the efficiency profile of the absolutely fair partition (Figure \ref{fig:fig3}B). The outcome of the actual partition is 13th in this sorted list (out of 433), i.e., it is located in the top 3\% of all the feasible outcomes. Therefore, we conclude that despite the fact that PSRM achieves the highest efficiency score among the four parties, there is no clear evidence for the hypothesis of gerrymandering in favor of PSRM or any other party.

Also, we sought to showcase that our approach for determining the fair districting is more appropriate and accurate than the one based on the comparison of the nation-wide results projected onto the actual districting and the electoral outcome computed by the rules of proportional voting. Parties for which the projected results are better than the proportional ones are sometimes deemed to gerrymander the districting map \cite{pasha2019gerrymandering}. Conversely, parties that under the actual districting win fewer districts than they would in the proportional system, are considered to be the ``victims'' of gerrymandering. We computed the electoral results of the nation-wide constituency by following the rules of the proportional system (the ``coefficient method'' used in 2019 elections \cite{codul}). We excluded from computation the votes coming from Transnistria and the diaspora and we also excluded the mandates corresponding to them. According to this procedure, PSRM would obtain 17 mandates, PDM - 12, ACUM - 13, and \c{S}OR - 4 (Figure \ref{fig:fig2}C). Note that in the single-member district system, \c{S}OR can never win 4 districts (since the theoretically maximal value of won districts is 3 for any districting map). Therefore, as in the actual districting \c{S}OR wins only 2 districts, it would be incorrectly marked as the ``victim'' of gerrymandering. In reality, as its efficiency score is 54.3\%, the actual map is neither the most optimistic, nor the most pessimistic one.

Finally, the real results of the single-district voting are very different from the nation-wide results projected onto the actual partition (Figure \ref{fig:uninominal}). In reality, PSRM won only 16 districts, PDM surprisingly obtained 17 districts, ACUM - 10, and \c{S}OR - 2 (Figure \ref{fig:fig2}D). These results suggest that many voters changed their nation-wide voting preferences to PDM in their local district. Several studies suggested that due to their financial and administrative resources, increased control over central and local authorities (before parliamentary elections PDM had 402 out of a total of 898 mayors \cite{mihaimogildea}), and all-encompassing media support \cite{odihr}, PDM was able to win a large number of mandates. We are not speculating on the explanation of the real results since our analysis is only concerned with the problem of gerrymandering. Although the actual districting was not very beneficial for PDM, it managed to achieve a striking performance in the single-member voting. PSRM's efficiency dropped down to only 7.0\%, ACUM's efficiency was as well poor - 15.7\%. \c{S}OR won its median number of districts achieving efficiency 54.3\%. PDM performed at the astonishing level of 99.4\%.

\subsection{Comparison with existing methods for detecting gerrymandering}

Detecting gerrymandering is a very important mathematical problem with applications to political sciences and multiple metrics for quantification of gerrymandering in districting maps have been proposed \cite{warrington2019comparison, duchin2018gerrymandering}. One of the most traditional metrics is the {\em efficiency gap} \cite{mcghee2014measuring, stephanopoulos2015partisan}. The efficiency gap is only defined in the case of bipartisan gerrymandering, and we are not aware of any of its variants that can be directly applicable to the case of many parties. Therefore, we contrast our approach with mean-median difference score (MM) \cite{warrington2019comparison} that can be compared with our approach in a straightforward way. MM score of a party is defined as the difference between its mean vote share and its median vote share across all the districts in a districting map. If the MM score is positive than the party is considered to be a victim of gerrymandering \cite{stephanopoulos2018measure}. 

\begin{figure}[ht]
  \centering
  \includegraphics[scale=0.35]{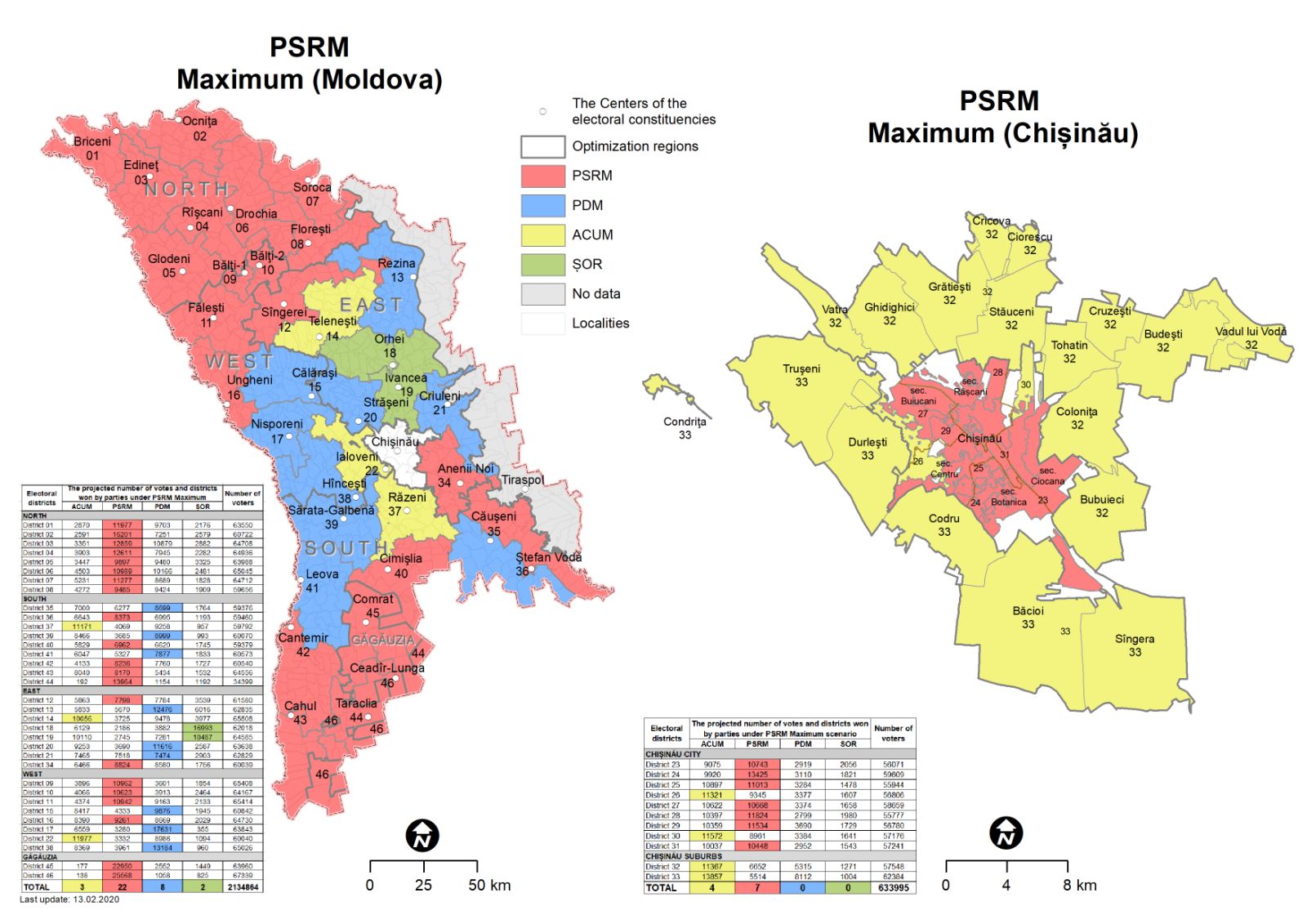}
  \caption{A maximum scenario for PSRM.}
  \label{fig:max-psrm}

  \vspace*{\floatsep}

  \includegraphics[scale=0.35]{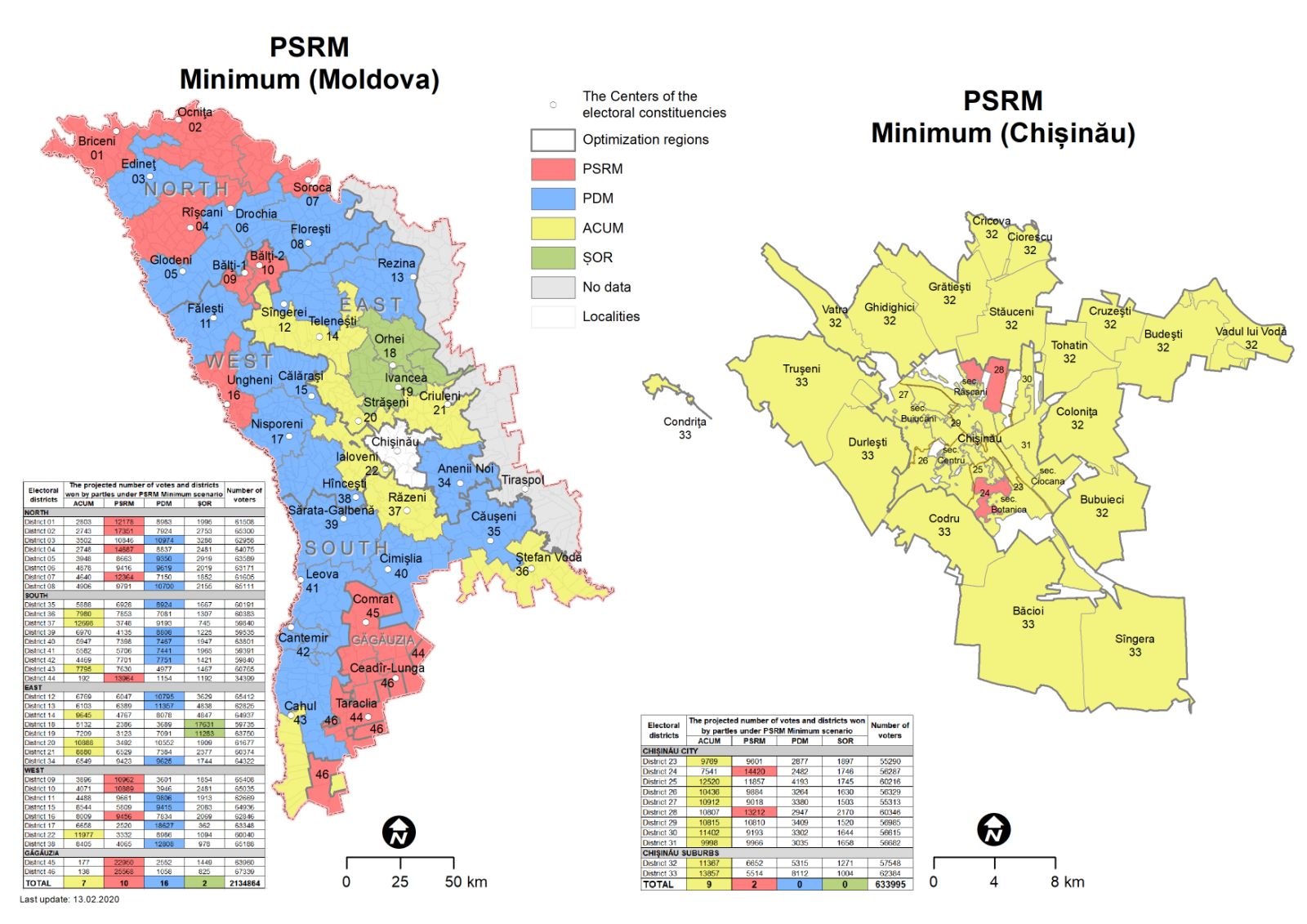}
  \caption{A minimum scenario for PSRM.}
  \label{fig:min-psrm}
\end{figure}

\begin{figure}[ht]
  \centering
  \includegraphics[scale=0.35]{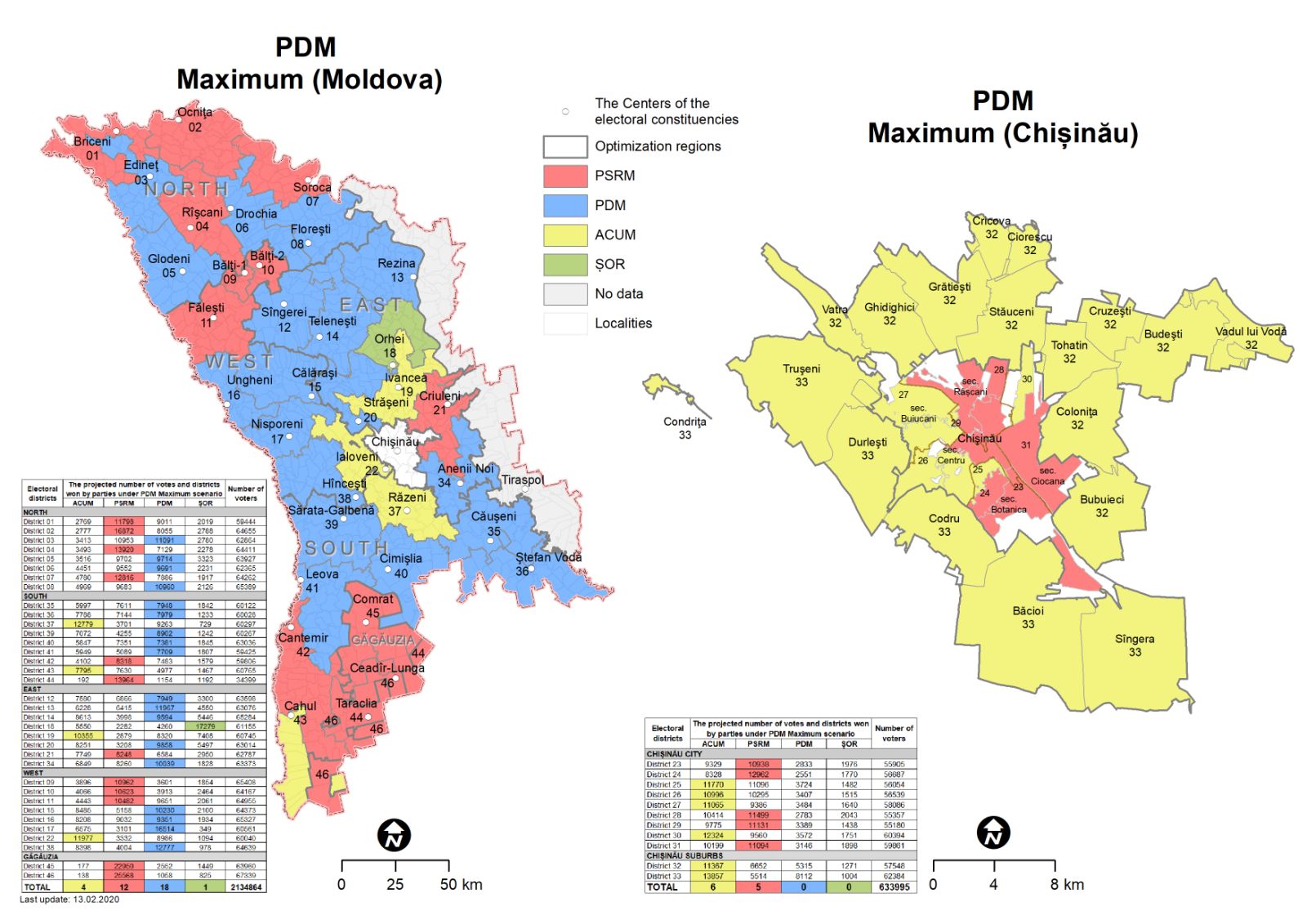}
  \caption{A maximum scenario for PDM.}
  \label{fig:max-pdm}

  \vspace*{\floatsep}

  \includegraphics[scale=0.35]{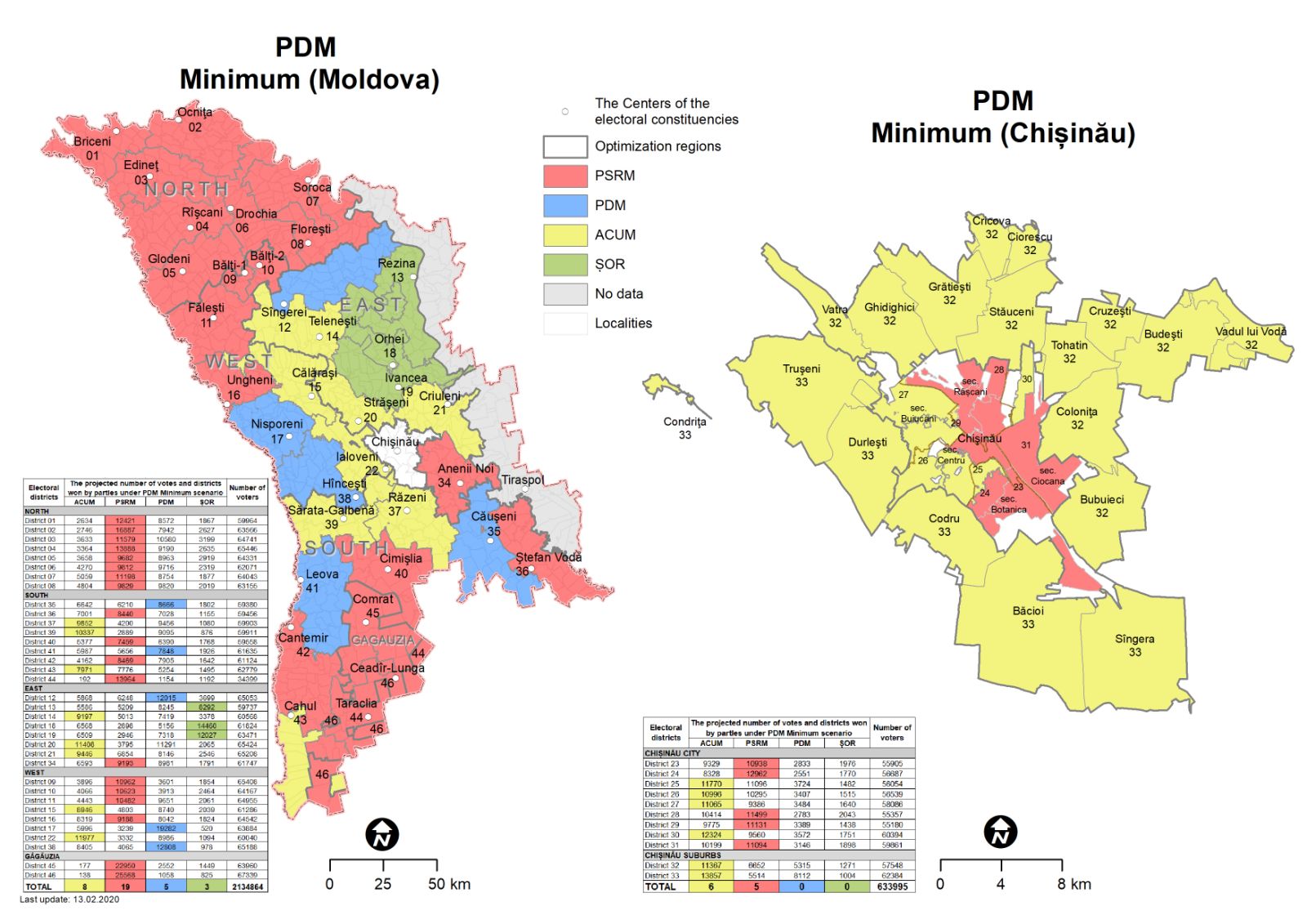}
  \caption{A minimum scenario for PDM.}
  \label{fig:min-pdm}
\end{figure}

\begin{figure}[ht]
  \centering
  \includegraphics[scale=0.35]{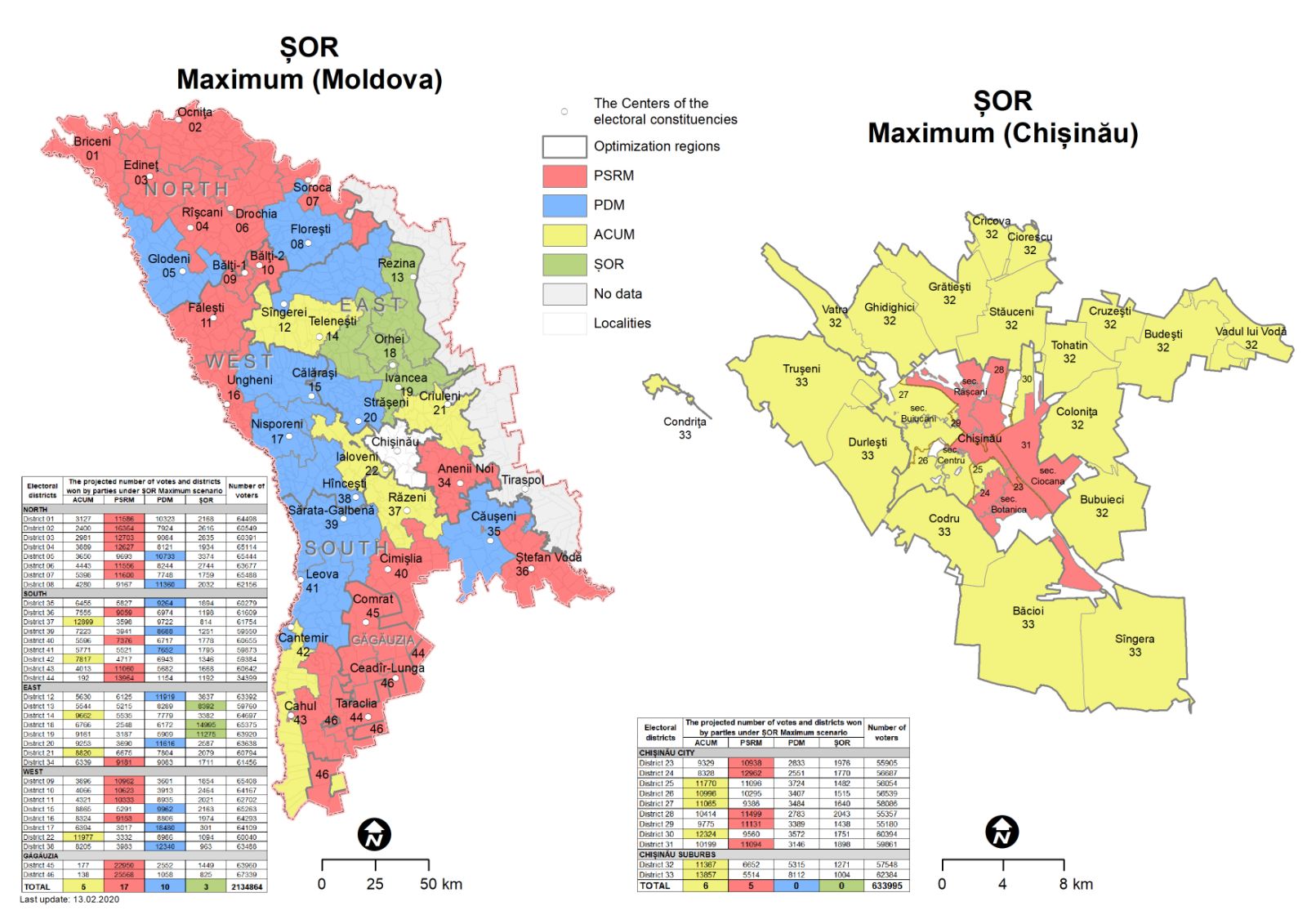}
  \caption{A maximum scenario for \c{S}OR.}
  \label{fig:max-sor}

  \vspace*{\floatsep}

  \includegraphics[scale=0.35]{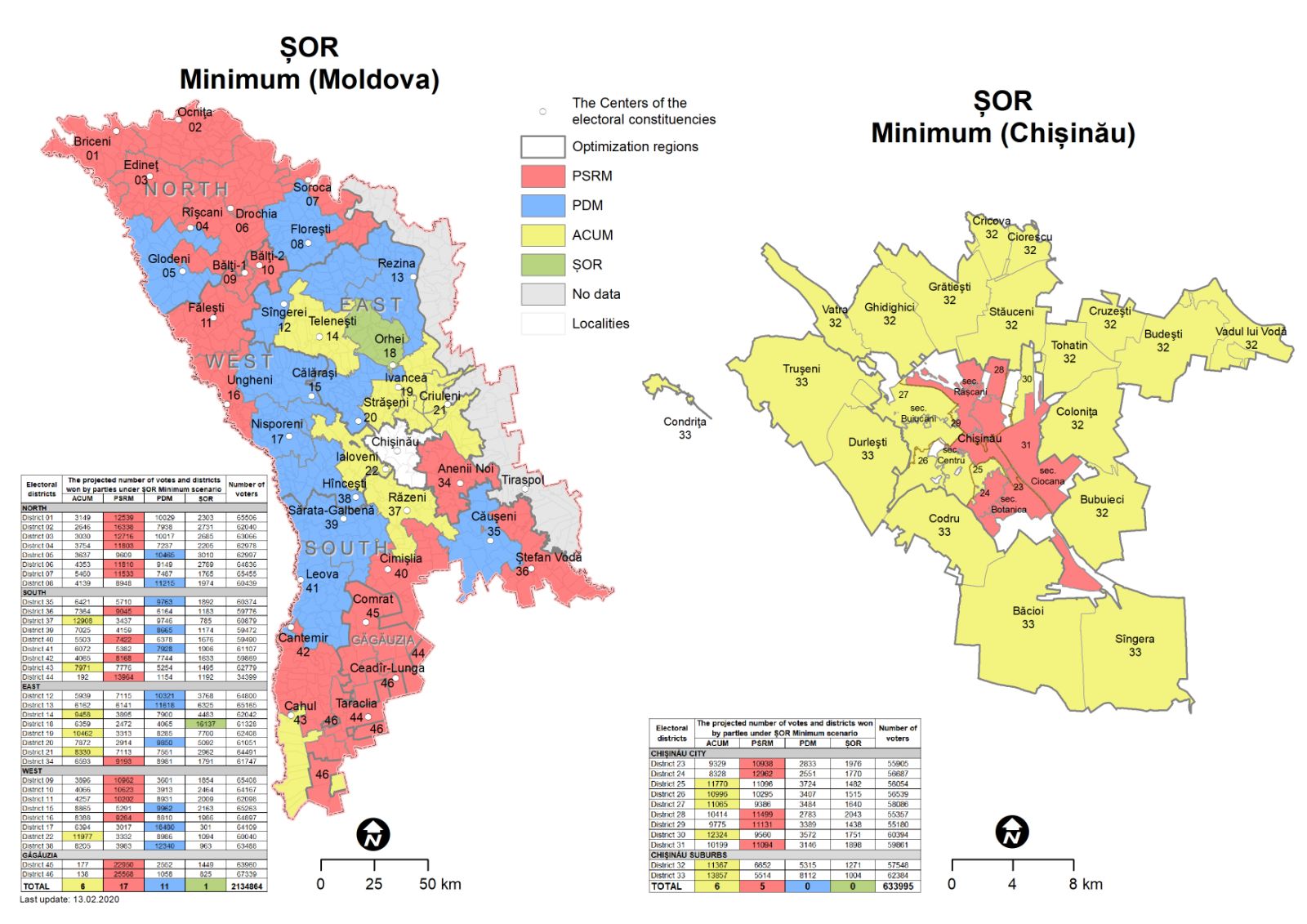}
  \caption{A minimum scenario for \c{S}OR.}
  \label{fig:min-sor}
\end{figure}

\begin{figure}[ht]
  \centering
  \includegraphics[scale=0.35]{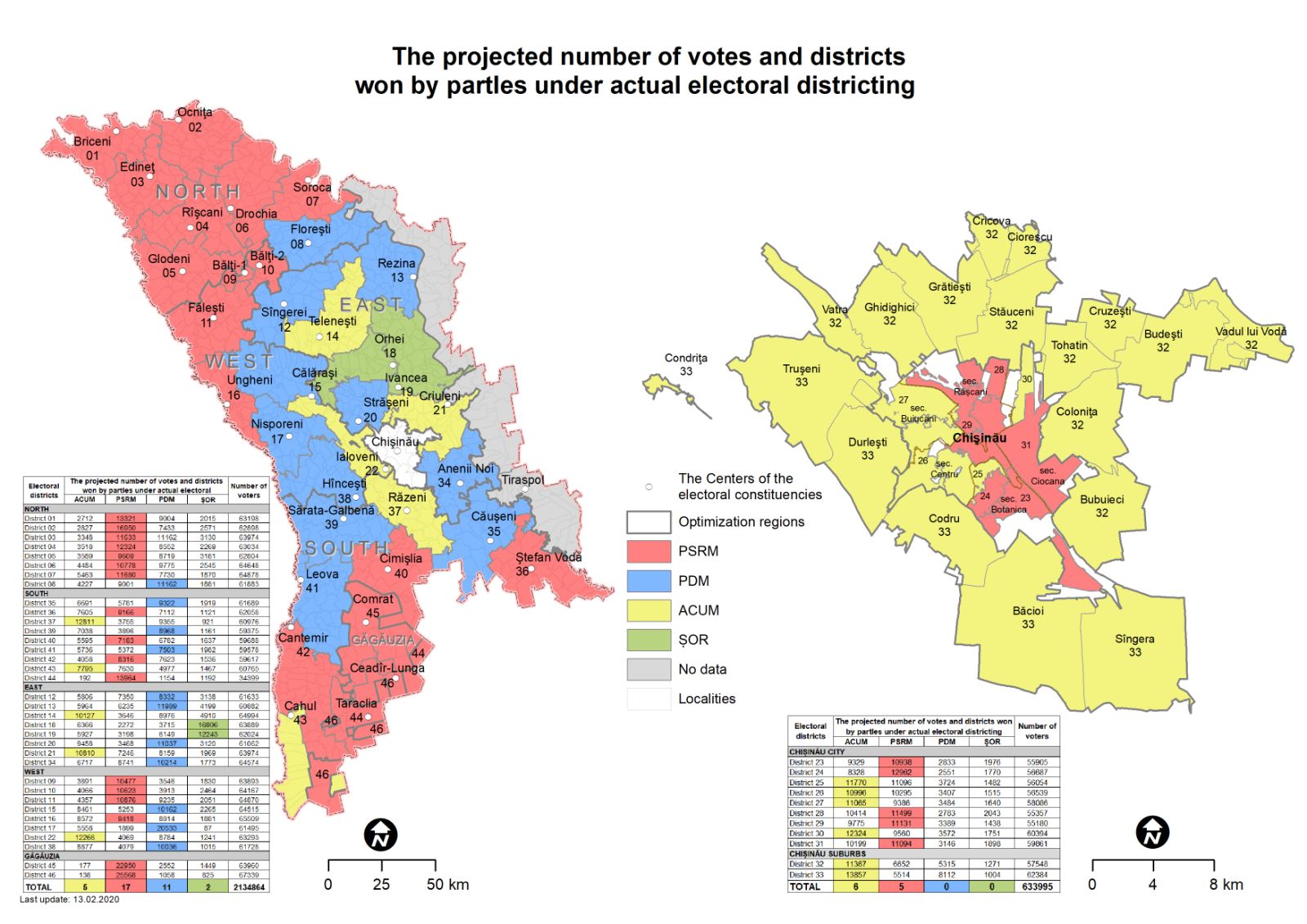}
  \caption{The results of the nation-wide constituency projected onto the actual districting map.}
  \label{fig:actual}

  \vspace*{\floatsep}

  \includegraphics[scale=0.35]{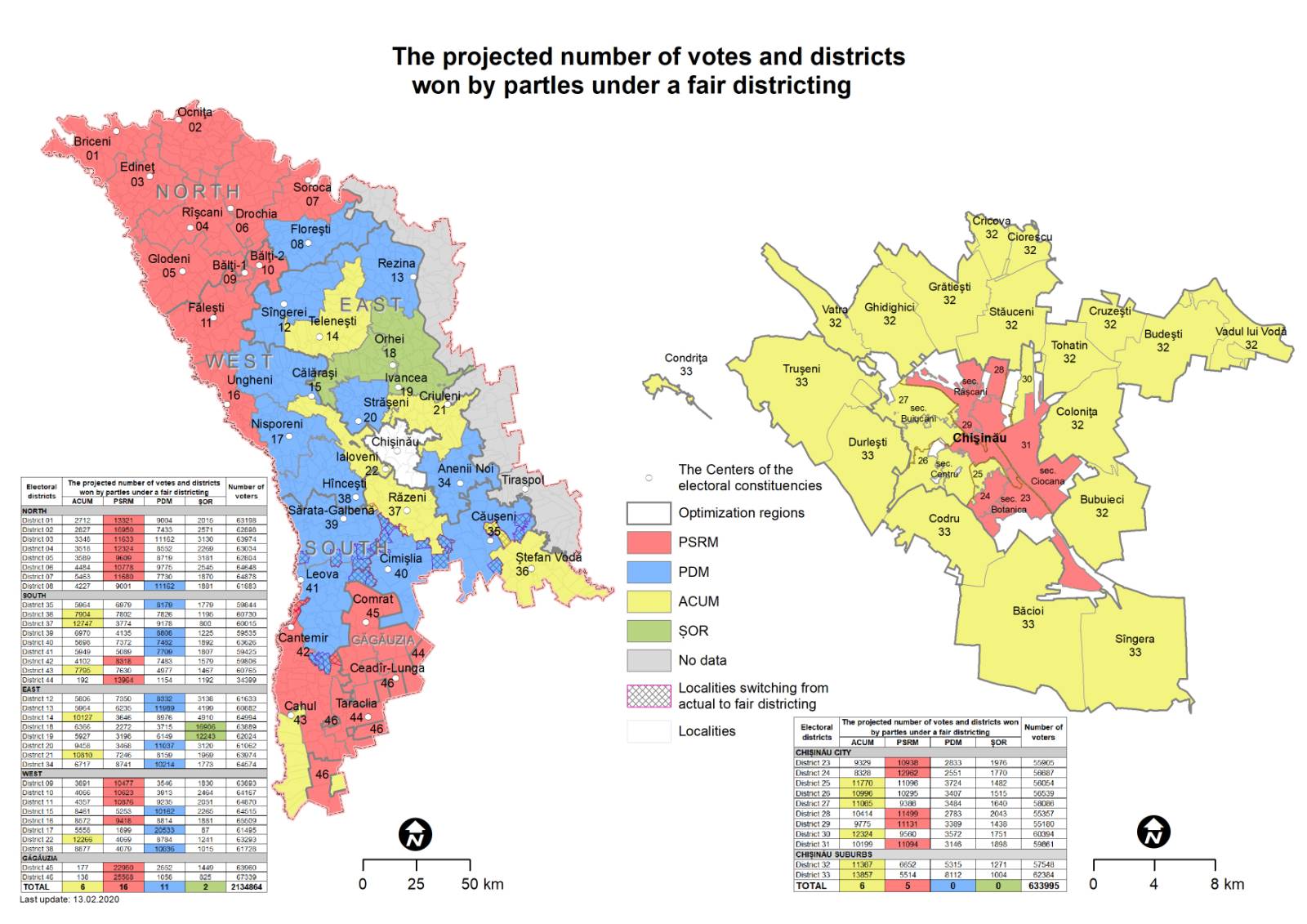}
  \caption{A fair districting map.}
  \label{fig:fair}
\end{figure}

\begin{figure}[ht]
  \centering
  \includegraphics[scale=0.35]{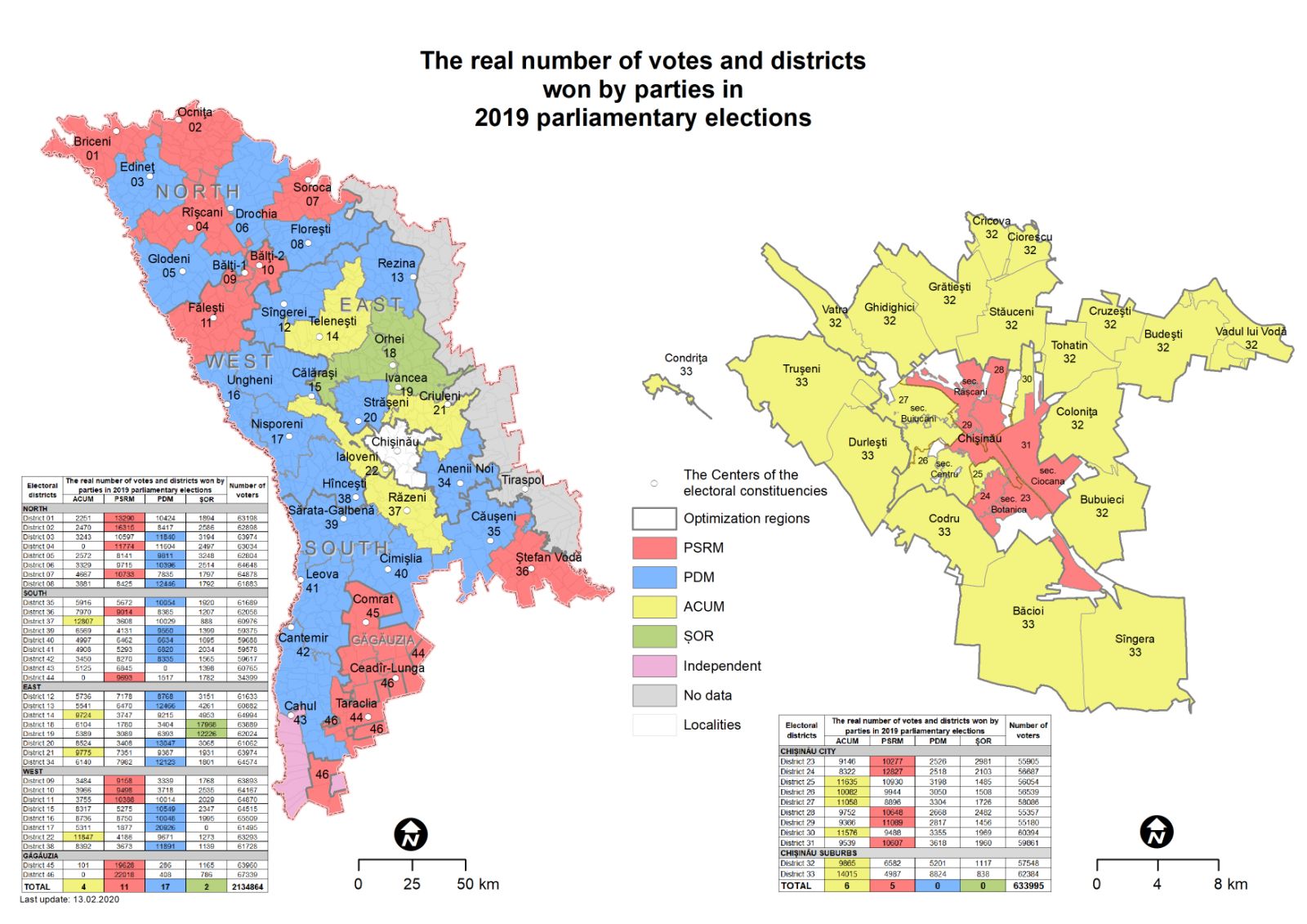}
  \caption{The actual results of the single-member district voting.}
  \label{fig:uninominal}

\end{figure}

We can illustrate on a counter-example that MM score is not an appropriate metric for detecting gerrymandering. Indeed, in the case of the districting map in Figure \ref{fig:max-psrm}, PSRM obtains its maximum possible number of mandates. The efficiency scores are the following: PSRM - 100, PDM - 5.9, ACUM - 0.6, and \c{S}OR - 54.3 (see Supplementary Table 2). However, when we compute the MM score for the same scenario we obtain the following results: PSRM - 0.009, PDM - --0.031, ACUM - 0.005, \c{S}OR - 0.022. The MM score of PSRM is positive; however, it is unreasonable to claim that it is a victim of gerrymandering.

\section{Conclusion}

We present a novel methodology for detecting gerrymandering and computing fair districting under parallel voting systems. Our methodology is based on identifying the set of all feasible electoral outcomes by first solving the Minimum and Maximum Gerrymandering Problems for finding the ranges of possible number of won districts, and then, identifying all the feasible outcomes by repeated solving of the Fixed Gerrymandering Problem. We applied our approach to the Moldovan parliamentary elections of 2019 and positively answered the intriguing question of whether the actual districting used in the single-member district election was relatively fair. Importantly, we also provided an example of the most equitable districting map that does not advantage any of the political parties.

We conclude with several caveats and future directions. First, due to the enormous search space of the gerrymandering problems, it is extremely challenging to estimate the frequency of each electoral outcome. In our approach, we considered that each outcome is equally likely. However, the adherents of statistical approaches to districting would hardly agree with this assumption.

Second, due to the same issue, in the analysis of Moldovan parliamentary elections, we restricted the search space to only geographically appropriate districting maps and we split the largest optimization regions into four parts. We acknowledge that some solutions could potentially be lost. However, we believe that skipping important solutions is highly unlikely.

Finally, an obvious future direction is to extend our approach to the case when several parties create different forms of alliances.

\bibliographystyle{unsrt}
\bibliography{references}

\begin{thebibliography}{10}

\bibitem{niemi1978theory}
Richard~G Niemi and John Deegan.
\newblock {A theory of political districting}.
\newblock {\em American Political Science Review}, 72(4):1304--1323, 1978.

\bibitem{williams1995political}
Justin~C Williams.
\newblock {Political redistricting: a review}.
\newblock {\em Papers in Regional Science}, 74(1):13--40, 1995.

\bibitem{garfinkel1970optimal}
Robert~S Garfinkel and George~L Nemhauser.
\newblock {Optimal political districting by implicit enumeration techniques}.
\newblock {\em Management Science}, 16(8):B--495, 1970.

\bibitem{lewyn1993limit}
Michael~E Lewyn.
\newblock {How to limit gerrymandering}.
\newblock {\em Fla. L. Rev.}, 45:403, 1993.

\bibitem{backstrom1977issues}
Charles Backstrom, Leonard Robins, and Scott Eller.
\newblock {Issues in gerrymandering: an exploratory measure of partisan
  gerrymandering applied to Minnesota}.
\newblock {\em Minn. L. Rev.}, 62:1121, 1977.

\bibitem{mcgann2016gerrymandering}
Anthony~J McGann, Charles~Anthony Smith, Michael Latner, and Alex Keena.
\newblock {\em {Gerrymandering in America: The House of Representatives, the
  Supreme Court, and the future of popular sovereignty}}.
\newblock Cambridge University Press, 2016.

\bibitem{malesky2009gerrymandering}
Edmund Malesky.
\newblock {Gerrymandering - Vietnamese style: escaping the partial reform
  equilibrium in a nondemocratic regime}.
\newblock {\em The Journal of Politics}, 71(1):132--159, 2009.

\bibitem{halas2017functionality}
Mari{\'a}n Hal{\'a}s and Pavel Klapka.
\newblock {Functionality versus gerrymandering and nationalism in
  administrative geography: lessons from Slovakia}.
\newblock {\em Regional Studies}, 51(10):1568--1579, 2017.

\bibitem{plener2018quantifying}
Benjamin Plener~Cover.
\newblock {Quantifying partisan gerrymandering: An evaluation of the efficiency
  gap proposal}.
\newblock {\em Stan. L. Rev.}, 70:1131, 2018.

\bibitem{veomett2018efficiency}
Ellen Veomett.
\newblock {Efficiency gap, voter turnout, and the efficiency principle}.
\newblock {\em Election Law Journal: Rules, Politics, and Policy},
  17(4):249--263, 2018.

\bibitem{tapp2019measuring}
Kristopher Tapp.
\newblock {Measuring Political Gerrymandering}.
\newblock {\em The American Mathematical Monthly}, 126(7):593--609, 2019.

\bibitem{hess1965nonpartisan}
Sidney~Wayne Hess, JB~Weaver, HJ~Siegfeldt, JN~Whelan, and PA~Zitlau.
\newblock {Nonpartisan political redistricting by computer}.
\newblock {\em Operations Research}, 13(6):998--1006, 1965.

\bibitem{kalcsics2015districting}
J{\"o}rg Kalcsics.
\newblock {Districting problems}.
\newblock In {\em Location science}, pages 595--622. Springer, 2015.

\bibitem{ricca2013political}
Federica Ricca, Andrea Scozzari, and Bruno Simeone.
\newblock {Political districting: from classical models to recent approaches}.
\newblock {\em Annals of Operations Research}, 204(1):271--299, 2013.

\bibitem{altman1997automation}
Micah Altman.
\newblock {Is automation the answer: The computational complexity of automated
  redistricting}.
\newblock {\em Rutgers Computer and Law Technology Journal}, 23, 1997.

\bibitem{nygreen1988european}
Bj{\o}rn Nygreen.
\newblock {European assembly constituencies for wales-comparing of methods for
  solving a political districting problem}.
\newblock {\em Mathematical Programming}, 42(1-3):159--169, 1988.

\bibitem{nemoto2003modelling}
T~Nemoto and K~Hotta.
\newblock {Modelling and solution of the problem of optimal electoral
  districting}.
\newblock {\em Communications of the OR Society of Japan}, 48:300--306, 2003.

\bibitem{li2007quadratic}
Zhenping Li, Rui-Sheng Wang, and Yong Wang.
\newblock {A quadratic programming model for political districting problem}.
\newblock In {\em Proceedings of the firsst international symposium on
  optimization and system biology (OSB). Bejing, China}, 2007.

\bibitem{hojati1996optimal}
Mehran Hojati.
\newblock {Optimal political districting}.
\newblock {\em Computers \& Operations Research}, 23(12):1147--1161, 1996.

\bibitem{bozkaya2003tabu}
Burcin Bozkaya, Erhan Erkut, and Gilbert Laporte.
\newblock {A tabu search heuristic and adaptive memory procedure for political
  districting}.
\newblock {\em European Journal of Operational Research}, 144(1):12--26, 2003.

\bibitem{ricca2008local}
Federica Ricca and Bruno Simeone.
\newblock {Local search algorithms for political districting}.
\newblock {\em European Journal of Operational Research}, 189(3):1409--1426,
  2008.

\bibitem{pasha2019gerrymandering}
{Gerrymandering 2.0: how were the uninominal constituencies in the {R}epublic
  of {M}oldova drawn?}
\newblock
  \url{https://watchdog.md/wp-content/uploads/2018/02/Gerry-Mandering-2.0-eng.pdf}.
\newblock Accessed: 2019-11-26.

\bibitem{fifield2015new}
Benjamin Fifield, Michael Higgins, Kosuke Imai, and Alexander Tarr.
\newblock {A new automated redistricting simulator using markov chain monte
  carlo}.
\newblock {\em Work. Pap., Princeton Univ., Princeton, NJ}, 2015.

\bibitem{chen2016evaluating}
Jowei Chen and David Cottrell.
\newblock {Evaluating partisan gains from Congressional gerrymandering: Using
  computer simulations to estimate the effect of gerrymandering in the US
  House}.
\newblock {\em Electoral Studies}, 44:329--340, 2016.

\bibitem{tam2016toward}
Wendy~K Tam~Cho and Yan~Y Liu.
\newblock {Toward a talismanic redistricting tool: A computational method for
  identifying extreme redistricting plans}.
\newblock {\em Election Law Journal}, 15(4):351--366, 2016.

\bibitem{borodin2018big}
Allan Borodin, Omer Lev, Nisarg Shah, and Tyrone Strangway.
\newblock {Big City vs. the Great Outdoors: Voter Distribution and How It
  Affects Gerrymandering.}
\newblock In {\em IJCAI}, pages 98--104, 2018.

\bibitem{538}
{The Gerrymandering Project}.
\newblock \url{https://fivethirtyeight.com/tag/the-gerrymandering-project/}.
\newblock Accessed: 2019-12-22.

\bibitem{harta}
{Harta circumscriptiilor uninominale}.
\newblock
  \url{https://gov.md/ro/content//harta-circumscriptiilor-uninominale-prezentata-spre-aprobare-guvernului}.

\bibitem{cecmd}
{Comisia Electorala Centrala a Republicii Moldova}.
\newblock \url{https://a.cec.md/ro/rezultate-alegeri-4287.html}.
\newblock Accessed: 2019-12-22.

\bibitem{codul}
{The Electoral Code of Moldova (12/30/2018 version)}.
\newblock \url{https://www.legis.md/cautare/getResults?doc_id=111378&lang=ro}.
\newblock Accessed: 2019-12-22.

\bibitem{mihaimogildea}
Mihai Mogildea and Diana Kralova.
\newblock {Between Theory and Practice: Possible outcomes of the Parliamentary
  Elections in Moldova under the mixed electoral system}.
\newblock
  \url{http://ipre.md/wp-content/uploads/2018/12/Research-Paper-Possible-scenarios-for-the-parliamentary-elections.pdf},
  2018.
\newblock Accessed: 2019-12-9.

\bibitem{odihr}
{ODIHR final report on Moldova's parliamentary elections of 24 February 2019}.
\newblock
  \url{http://www.epgencms.europarl.europa.eu/cmsdata/upload/3722ab72-5ca4-4de2-aaa3-e04a62c2799f/Moldove-parliamentary-elections__24-Feb-2019-OSCE_final_report.pdf},
  2019.
\newblock Accessed: 2019-12-9.

\bibitem{warrington2019comparison}
Gregory~S Warrington.
\newblock A comparison of partisan-gerrymandering measures.
\newblock {\em Election Law Journal: Rules, Politics, and Policy},
  18(3):262--281, 2019.

\bibitem{duchin2018gerrymandering}
Moon Duchin.
\newblock Gerrymandering metrics: How to measure? what's the baseline?
\newblock {\em arXiv preprint arXiv:1801.02064}, 2018.

\bibitem{mcghee2014measuring}
Eric McGhee.
\newblock Measuring partisan bias in single-member district electoral systems.
\newblock {\em Legislative Studies Quarterly}, 39(1):55--85, 2014.

\bibitem{stephanopoulos2015partisan}
Nicholas~O Stephanopoulos and Eric~M McGhee.
\newblock Partisan gerrymandering and the efficiency gap.
\newblock {\em U. Chi. L. Rev.}, 82:831, 2015.

\bibitem{stephanopoulos2018measure}
Nicholas~O Stephanopoulos and Eric~M McGhee.
\newblock The measure of a metric: The debate over quantifying partisan
  gerrymandering.
\newblock {\em Stan. L. Rev.}, 70:1503, 2018.

\end{thebibliography}

\newpage

\end{document}